\documentclass[twocolumn,prl,superscriptaddress,showpacs]{revtex4}
\usepackage{graphicx,amsmath,amssymb,color}
\usepackage{dcolumn}
\usepackage{bm}
\usepackage[ansinew]{inputenc}

\begin{document}

\title {Carrier dynamics in epitaxial graphene close to the Dirac point}

\author{S. Winnerl}
\email{s.winnerl@hzdr.de}
\affiliation{%
Helmholtz-Zentrum Dresden-Rossendorf, P.O. Box 510119, 01314 Dresden, Germany
}%
\author{M. Orlita}%
\affiliation{%
Grenoble High Magnetic Field Laboratory, CNRS-UJF-UPS-INSA F-38042 Grenoble Cedex 09, France
}%
\author{P. Plochocka}
\affiliation{%
Grenoble High Magnetic Field Laboratory, CNRS-UJF-UPS-INSA F-38042 Grenoble Cedex 09, France
}%
\author{P. Kossacki}
\affiliation{%
Grenoble High Magnetic Field Laboratory, CNRS-UJF-UPS-INSA F-38042 Grenoble Cedex 09, France
}%
\author{M. Potemski}
\affiliation{%
Grenoble High Magnetic Field Laboratory, CNRS-UJF-UPS-INSA F-38042 Grenoble Cedex 09, France
}%
\author{T. Winzer}
\affiliation{%
Technische Universität Berlin, Hardenbergstr. 36 10623 Berlin, Germany
}%
\author{E. Malic}
\affiliation{%
Technische Universität Berlin, Hardenbergstr. 36 10623 Berlin, Germany
}%
\author{A. Knorr}
\affiliation{%
Technische Universität Berlin, Hardenbergstr. 36 10623 Berlin, Germany
}%
\author{M. Sprinkle}
\affiliation{%
Georgia Institute of Technology, Atlanta, Georgia 30332, USA
}%
\author{C. Berger}
\affiliation{%
Georgia Institute of Technology, Atlanta, Georgia 30332, USA
}%
\author{W.A. de Heer}
\affiliation{%
Georgia Institute of Technology, Atlanta, Georgia 30332, USA
}%
\author{H. Schneider}
\affiliation{%
Helmholtz-Zentrum Dresden-Rossendorf, P.O. Box 510119, 01314 Dresden, Germany
}%
\author{M. Helm}
\affiliation{%
Helmholtz-Zentrum Dresden-Rossendorf, P.O. Box 510119, 01314 Dresden, Germany
}%

\date{\today}

\begin{abstract}
We study the carrier dynamics in epitaxially grown graphene in the range of photon energies from 10 - 250 meV. The experiments complemented by microscopic modeling reveal that the carrier relaxation is significantly slowed down as the photon energy is tuned to values below the optical phonon frequency, however, owing to the presence of hot carriers, optical phonon emission is still the predominant relaxation process. For photon energies about twice the value of the Fermi energy, a transition from pump-induced transmission to pump-induced absorption occurs due to the interplay of interband and intraband processes.
\end{abstract}

\pacs{78.67.Wj, 63.22.Rc, 78.47.jd}
\maketitle

%



Graphene, consisting of a single atomic layer of carbon atoms in a hexagonal lattice, exhibits a unique band structure with zero energy gap and linear energy dispersion, the so-called Dirac cone. The band structure gives rise to several remarkable properties, some of which are highly attractive for novel optoelectronic and photonic devices \cite{Bonaccorso2010}. Of key importance are the dynamics of electronic relaxation, in particular due to the interaction with the phonon system. During the last years many insights into the relaxation dynamics have been obtained from single-color \cite{Dawlaty2008, Plochocka2009, Breusing2009, Wang2010} and two-color \cite{Sun2008, Newson2009, George2008, Choi2009, Sun2010} pump-probe experiments. Thermalization of the nonequilibrium electron distribution via electron-electron scattering on a sub-100 fs timescale and efficient scattering via optical phonons on a 100 fs - few ps timescale have been identified. 
Common to previous pump-probe experiments is an excitation energy of $\sim$1.5 eV, i.e. high above the Dirac point. Also most graphene photonic applications demonstrated so far involve near infrared or visible light \cite{Xia2009, Xing2010}. The nature of graphene as a gapless material with constant absorption in the range, where the band structure is well described by Dirac cones, suggests to expand the studies into the mid infrared and terahertz range. 
In particular, it is crucial to obtain thorough insights into the relaxation dynamics in the range of the optical phonon energy and below, i.e., close to the Dirac point, where Coulomb as well as optical and acoustic phonon processes can be significant and were both interband and intraband processes are relevant. Here graphene serves as a model system to understand the relevance of electron-electron and electron-phonon interaction for both intraband and interband relaxation in materials with small or vanishing energy gap.

In this Letter, we study the carrier dynamics close to the Dirac point by varying the excitation energy $E$ by more than an order of magnitude (245 meV - 10 meV).
An optical phonon bottleneck is observed in the range $E$ = 245 - 30 meV, with decay times increasing from sub-ps to several 100 ps. Microscopic calculations based on
the density matrix formalism explain these time constants by optical phonon scattering and additionally reveal contributions due to Coulomb- and acoustic phonon-induced processes.
For $E\:{<}$ 30 meV, a striking and unexpected sign reversal of the pump-probe signal is found. 
The involved processes are understood by a simple model based on the dynamic conductivity including \emph{intra-} and \emph{interband} absorption.
%

We study multilayer graphene containing $\sim$70 layers grown by thermal decomposition on the carbon terminated surface of 4H-SiC \cite{Berger2004}. It has been shown that such samples behave graphene-like rather than graphitic due to rotational stacking of layers which preserves the electronic symmetry of a single graphene layer \cite{Hass2008}. The graphene-like nature of our samples has been verified by Raman spectroscopy \cite{Ferrari2006} and magneto-spectroscopy \cite{Sadowski2006, Orlita2008}. In contrast to a recent two-color experiment on similar samples, where parts of the signal could be selectively attributed to the few highly doped graphene layers at the interface \cite{Sun2010}, our single-color study with much lower excitation energies investigates the relaxation dynamics in the large number of almost intrinsic graphene layers on top of the interface layers. Since our photon energies are below twice the value of the Fermi energy of the interface layers, no interband transitions in these layers are excited. The free-electron laser FELBE serves as a source for tunable picosecond pulses of infrared radiation \cite{Lehnert}. FELBE operates in a wavelength range from $4 - 280\:\mu m$ (photon energy $E$ = 310 - 4 meV) with a repetition rate of 13 MHz. In the experiments (performed at $E$ = 245 meV, 72 meV, 51 meV, 30 meV, 20 meV, 18 meV, 14 meV, 8 meV \cite{reststrahlenband}) the pump beam was mechanically chopped, the polarization of the probe beam was rotated by 90° with respect to the pump beam and a polarization analyzer was placed in front of the detector in order to minimize scattered pump radiation. Mercury-cadmium-telluride detectors optimized for the respective wavelength were used for the three highest photon energies. For the lower photon energies a liquid He cooled extrinsic Ge:Ga detector was employed.

\begin{figure}
\includegraphics[width=8cm]{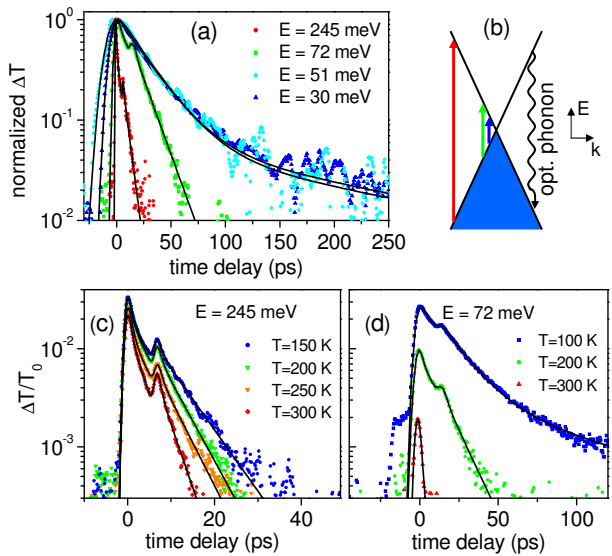}
\caption{\label{fig:epsart} (Color online) Experiment: Normalized pump-induced transmission for different photon energies $E$ (a) measured at 10 K. Illustration of the optical pumping (arrows in colors like in part (a)) in comparison with relaxation via optical phonons (b). Temperature dependence of the pump-induced transmission for E = 245 meV (c) and $E$ = 72 meV (d). The symbols are experimental data, the lines are fits taking into account the laser pulse duration and a bi-exponential decay.}
\end{figure}

\begin{table}
\caption{\label{tab:table5}Full width at half maximum (FWHM) of the FEL pulse and the decay components of the pump-probe signals measured at $T$ = 10 K. For small photon energies, $\tau_{1}$ is below the experimental time resolution.}
\begin{ruledtabular}
\begin{tabular}{ccccc}
E(meV)& $\tau_\text{pulse}\:$ (ps) & $\tau_\text{1}\:$ (ps) & $\tau_\text{2}\:$ (ps) & $\tau_\text{3}\:$ (ps)\\
\hline
245 & 0.7 & 0.5$\:\pm\:$0.1 & 5.2$\:\pm\:$0.2 & within noise \\
72 & 3 & - & 14.5$\:\pm\:$1 & 300$\:\pm\:$50\\
51 & 11 & - & 25$\:\pm\:$2 & 300$\:\pm\:$50\\
30 & 7 & - & 25$\:\pm\:$2 & 300$\:\pm\:$50\\
\end{tabular}
\end{ruledtabular}
\end{table}
In Fig. 1(a) the pump-induced transmission change obtained for different photon energies down to 30 meV is shown. The lattice temperature was 10 K. The different rise times correspond to different durations of the FELBE pulses (see Tab. 1). The secondary peaks observed in the signals for the two highest photon energies are artifacts due to a second pump pulse which is caused by internal reflections in beamline windows used for the respective wavelengths. To extract relaxation times the experimental data were fitted taking into account a Gaussian excitation pulse and a bi-exponential decay. The resulting time constants are summarized in Tab. 1: For $E$ = 245 meV, we observe a fast (0.5 ps) and a slower (5.2 ps) relaxation time. The observation of two timescales is consistent with previous results obtained in experimental studies in the near infrared showing a fast component ranging from 70 - 120 fs and a slower phonon-induced component in the range of 0.4 - 2.5 ps \cite{Dawlaty2008, Plochocka2009, Wang2010}. Our room temperature value $\tau_{2}$ = 3.4 $\pm$ 1 ps obtained for $E$ = 245 meV is only slightly longer, thus pointing to an optical-phonon related mechanism. Note that at this excitation energy the relaxation is due to efficient \emph{interband} scattering on optical phonons of an energy of $\sim$200 meV because direct \emph{intraband} relaxation is strongly suppressed 
(cf. Fig. 1(b)). In contrast, for near-infrared excitation, \emph{intraband} relaxation is the dominating relaxation channel. As the excitation energy is reduced to values below the optical phonon energy, $\tau_{2}$ is strongly increased to 14 - 25 ps. Additionally an even longer timescale of $\tau_{3}$ = 300 $\pm$ 50 ps is observed, which was buried in the noise floor for $E$ = 245 meV. Furthermore, we find an interesting temperature dependence: While for $E$ = 245 meV both the maximum transmission change and the relaxation times depend only weakly on temperature, both quantities drop sharply with increasing temperature for E = 72 meV (cf. Fig. 1 (c)-(d)). At room temperature the relaxation for E = 72 meV is equally fast as for E = 245 meV within the measurement accuracy.

To understand the involved relaxation dynamics more thoroughly, we use a microscopic theory \cite{Winzer2010} to calculate the relaxation dynamics of non-equilibrium carriers resolved in time, momentum, and angle. Within the density matrix formalism, we derive the graphene Bloch-equations \cite{Lindberg1988, Winzer2010, Butscher2007} describing the momentum dependent carrier density $\rho_{\lambda{\bf k}}$ (in the valence and conduction band), the microscopic polarization, and the phonon occupation number taking into account both optical and accoustic phonons at the $\Gamma$ and $K$ points. The Coulomb and the carrier-phonon interactions are considered using second order Born-Markov approximation resulting in a microscopic Boltzmann-like scattering equation with time and momentum dependent scattering rates and explicitly including Pauli blocking terms. To compare our microscopic results with the experiment, we approximate the differential transmission signal by the change of the carrier density in the optically excited state $\Delta\rho_{k_0}$ \cite{Breusing2011}. 
\begin{figure}
\includegraphics[width=8cm]{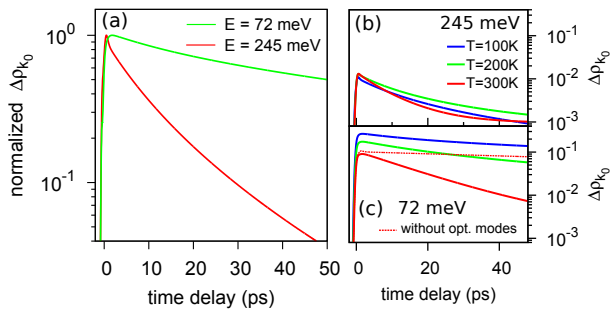}
\caption{\label{fig:epsart} (Color online) Theory: Normalized change of the carrier density of optically excited states for different photon energies at 10 K (a). Temperature dependence of the relaxation dynamics at $E$ =245 meV (b) and $E$ = 72 meV (c) for $T$ = 100 K, 200 K, 300 K. The dashed line is calculated without optical phonon modes illustrating their importance even at the low energy.}
\end{figure}
\begin{figure}
\includegraphics[width=8cm]{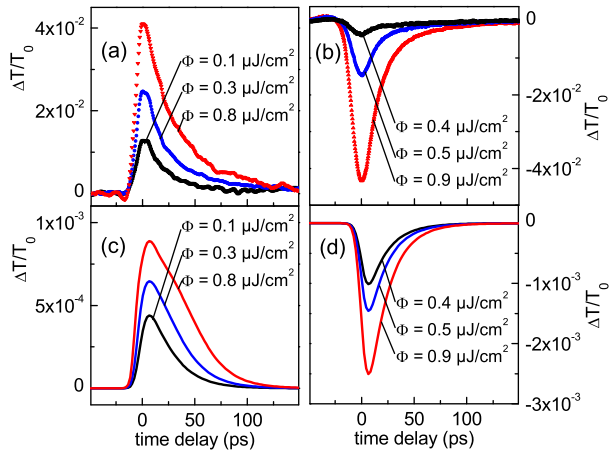}
\caption{\label{fig:epsart} (Color online) Experimental pump-induced transmission change for three different fluences for the photon energy 30 meV (a) and 20 meV (b). Calculated pump-induced transmission of a single graphene layer for photon energies of 30 meV (c) and 20 meV (d) obtained with the simple model based on electron heating described in the text.}
\end{figure}
In accord with the experiment we find that the relaxation is significantly slowed down as the photon energy is reduced below the optical phonon energy (Fig. 2(a)). Also the calculated temperature dependence (Fig.2(b)-(c)) confirms the experimental observations (Fig.1(c)-(d)). However, in contrast straight-forward expectations \cite{Ryzhii2007}, optical phonons are still the dominant process for cooling the electronic system, even if the excitation energy is below the optical phonon energy. This is illustrated in Fig. 2(c), where the curve for $T$ = 300 K is also calculated with the optical-phonon contribution switched off. The reason for this is the strong carrier-carrier interaction in graphene which provides a hot carrier distribution on sub-ps timescales, allowing hot electrons to scatter efficiently with optical phonons. This is energetically possible because some fraction of the hot electrons have energies exceeding the the optical-phonon energy.
The scattering by acoustic phonons is slower and leads to a decay time $\tau_{3}\approx$ 300 ps, which is also consistent with the experiment. For $E$ = 245 meV the calculation reveals another feature seen in the experiment, namely an initial fast time constant $\tau_{1}$. The microscopic calculation allows us to identify this contribution as Auger-type processes: the predominant impact excitation redistributes the energy of the excitation pulse to an increasing number of charge carriers leading to a fast cooling of the electronic system. Unlike the initial thermalization process, where only conduction band electrons are involved, the impact ionization transfers electrons form the valence band into the conduction band \cite{Winzer2010}. While the microscopic calculations give an excellent qualitative agreement with experimental results, the quantitative values at low temperature for $\tau_{1}$ and $\tau_{2}$ are by about a factor of 2 larger in the theory, which could be due to neglected non-Markov channels or scattering via defects. The experimentally observed effect of increased temperature on both the relaxation time and the maximum transmission change is stronger (especially for E = 72 meV) than in the calculation. One possible reason for this is the neglected momentum dependence in the electron-phonon coupling matrix elements and the phonon dispersion, which is here considered linear for acoustic phonons and constant for optical phonons. Furthermore the role of the SiC substrate is only included via a global dielectric screening constant. The effects discussed here are expected to result in modified relaxation times especially at elevated temperatures, where they become more prominent due to the broader thermal distribution.

Now we turn to discussing the carrier dynamics in the vicinity of 2$E_{f}$ of the almost intrinsic graphene layers. As $E$ is decreased from 30 to 20 meV, a striking change in the sign of the pump-probe signals is observed (Fig. 3 (a) and (b). The positive pump-probe signal, i.e. pump-induced transmission, is caused by the bleaching of the interband transition (cf. right inset of Fig. 4) similar to the cases discussed above. If the photon energy becomes smaller than twice the value of the Fermi energy, this process is not possible anymore. However, the carrier distribution can be heated via free-carrier absorption. As a result, states for which interband absorption is possible get populated with electrons (or equivalenty, depleted of holes) as indicated in the left inset of Fig.4 \cite{Doping}. \emph{Hence, pumping in this regime results in induced absorption}. For a quantitative description, we consider the real part of the dynamic conductivity \cite{Mikhailov2007, Choi2009}
\begin{align}
Re(\sigma(\omega))&=\frac{8\sigma_{0}\tau k_{B}T_{el}}{\pi \hbar(\omega \tau +1)}ln\left(2cosh(-E_{f}/2 k_{B} T_{el})\right)+\nonumber\\
&\frac{\sigma_{0}}{2}\left[tanh\left(\frac{\hbar \omega+2 E_{f}}{4 k_{B} T_{el}}\right)+tanh\left(\frac{\hbar \omega-2 E_{f}}{4 k_{B} T_{el}}\right)\right].
\end{align}
%
Here  $\sigma_{0} = e^{2}/4\hbar$ is the "universal value" of the dynamic conductivity, $T_{el}$ the carrier temperature and $\tau$ the momentum relaxation time. The first summand of eq. (1) indicates intraband Drude absorption, the second one interband absorption.  In Fig. 4 the dynamic conductivity is plotted for $E_{f}$ = -13 meV and $\tau$ = 300 fs. This value for the momentum relaxation time has been deduced from the linewidth in magneto-spectroscopy experiments \cite{Orlita2008}. Even at 10 K the Drude peak at low energies merges with the tail of the interband absorption, i.e. there is finite absorption at all energies. The arrows indicate the increase (decrease) of the absorption with increasing temperature for the photon energy 20 meV (30 meV). 
\begin{figure}
\includegraphics[width=75mm]{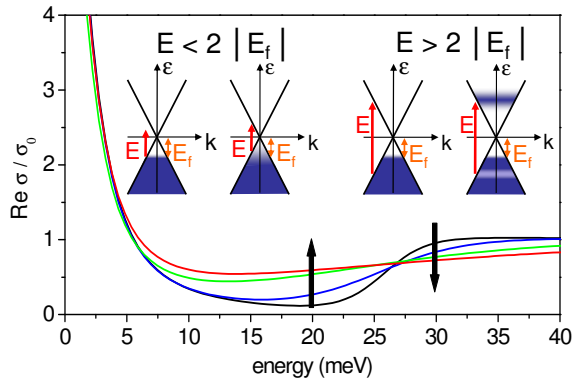}
\caption{\label{fig:epsart} (Color online) The dynamic conductivity for different temperatures (black: 10 K, blue: 20 K, green: 50 K and red 100 K). The inset sketches the carrier distribution before and after the pump beam for $E \;{<} \;2 |E_{f}|$ and $E \;{>}\; 2 |E_{f}|$.}
\end{figure}
For comparison with the experimental data of Fig. 3(a)-(b), we consider heating and relaxation of the electron distribution via the equation $dT_{el}/dt = P(t)/c_{p}-(T_{el}-T_{latt})/\tau_{2}$, where  $P(t) = I(t) Re(\sigma(T_{el}))/\epsilon_{0}c$ is the absorbed power per area ($I$, laser intensity; $c$, speed of light), $c_{p}$ the specific heat \cite{cp}, and $\tau_{2}$ = 25 ps a phenomenological energy relaxation time (see Tab.1). This simple model cannot account for the full dynamics; however, it contains the most important processes, since, at the timescales relevant here, a thermalized carrier distribution can be expected. The calculated transmission changes for a single graphene layer are plotted in Fig. 3 (c) and (d). The model captures the experimentally observed sign reversal. The agreement with the experiment is best for $|E_{f}| = 13 \pm 2$ meV. From magneto-spectroscopy experiments of samples grown with the same method $|E_{f}|$ = 8 meV has been concluded \cite{Orlita2008}. 
We note that the interplay of interband and intraband processes has been suggested recently to explain a change in sign of the pump-probe signal as pump fluence is varied in a UV-pump near-infrared-probe experiment \cite{Malard2011}. 

In summary, we have studied the relaxation dynamics of carriers in graphene close to the Dirac point. A slowing down of relaxation has been observed for energies below the optical phonon energy, but optical phonons still play an important role, as shown through microscopic calculations. The theoretical description of the results shows that the interplay of electron-electron and electron-phonon scattering produce new and efficient relaxation channels in situations when both intraband and interband relaxation are allowed, as they are in many narrow or vanishing gap systems. Most remarkably, the sign of the pump-probe signal is inverted, as the photon energy is varied across the value of 2$E_{f}$. This behavior can be understood by the interplay between inter- and intraband absorption and, besides being a fascinating behavior by itself, it allows a rather accurate determination of the Fermi energy. Furthermore the effect can be the basis of innovative THz optical switches and active filters.

We thank F. Peter, M. Mittendorff, O. Drachenko, D. Stehr and W. Seidel for friendly collaboration. Support from the German Science Foundation DFG in the framework of the Priority Program 1459 "Graphene" and from GACR via grants P204/10/1020 and GRA/10/E006 is gratefully acknowledged. The research at the free-electron laser FELBE was supported by the European Community's Seventh Framework Programme (FP7/2007-2013) under grant agreement n.°226716. We are grateful to P. Michel and the FELBE team for their dedicated support. 


\begin{thebibliography}{99}
\bibitem{reststrahlenband}
No experiments could be performed in the range from 100 - 200 meV due to the reststrahlenband of the SiC substrate.
\bibitem{Doping}
This is for p-type doping. For n-type, states in the the conduction band get depletd of electrons. The type of doping can not be determined by our experiments and does not affect the interpretation.
\end{thebibliography}
\end{document}